\documentclass[11pt]{article}

\usepackage[dvips]{graphicx}
\usepackage{amsmath}
\usepackage{amsfonts}
\usepackage{amssymb}

\newcommand{\beq}{\begin{equation}}
\newcommand{\eeq}{\end{equation}}
\newcommand{\R}{{\mathbb{R}}}
\newcommand{\C}{{\mathbb{C}}}
\newcommand{\SSSS}{{\mathbb{S}}}
\newcommand{\rmd}{\textrm{d}}
\newcommand{\rmi}{\textrm{i}}
\newcommand{\Imag}{\textrm{Im}\,}
\newcommand{\Real}{\textrm{Re}\,}
\newcommand{\Cons}{\,\textrm{Const.}\,}
\newcommand{\bj}{\mathbf{j}}
\newcommand{\bR}{\mathbf{R}}
\newcommand{\cU}{{\cal U}}
\newcommand{\sds}{\strut\displaystyle}
\newcommand{\smallspace}{\vspace{2.5ex}}

\begin{document}

\title{\bf Unified Scheme for Describing Time Delay and Time Advance in the 
Interpolation of Rotational Bands of Resonances}

\author{\vspace{5pt} Enrico De Micheli$^{*}$ and Giovanni Alberto Viano$^{\dag}$ \\
$^{*}$\small{IBF - Consiglio Nazionale delle Ricerche}\\[-5pt]
\small{Via De Marini, 6 - 16149 Genova, Italy. \vspace{8pt}} \\
$^{\dag}$\small{Dipartimento di Fisica - Universit\`a di Genova} \\[-5pt]
\small{Istituto Nazionale di Fisica Nucleare - sez. di Genova} \\[-5pt]
\small{Via Dodecaneso, 33 - 16146 Genova, Italy.}
}

\date{}

\maketitle

\begin{abstract}
In this paper we show how rotational bands of resonances can be described 
by using trajectories of poles of the scattering amplitude in the complex 
angular momentum plane: each band of resonances is represented by the 
evolution of a single pole lying in the first quadrant of the plane. 
The main result of the paper consists in showing that also the antiresonances 
(or echoes) can be described by trajectories of the scattering amplitude poles, 
instead of using the hard--sphere potential scattering as prescribed by the 
classical Breit--Wigner theory. The antiresonance poles lie in the fourth 
quadrant of the complex angular momentum plane, and are associated with 
non--local potentials which take into account the exchange forces; 
it derives a clear--cut separation between resonance and antiresonance poles.
The evolution of these latter poles describes the passage from quantum to 
semi--classical physics. The theory is tested on the rotational band produced 
by $\alpha$--$\alpha$ elastic scattering and on the hadronic rotational bands 
in $\pi^+$--p elastic scattering.
\end{abstract}


\maketitle

\section{Introduction}
\label{se:introduction} 
The standard partial--wave expansion of the scattering amplitude can be regarded 
as a factorization which splits the scattering amplitude into two factors: 
the partial--wave amplitudes $a_\ell(E)$ ($E$ being the center of mass energy, 
$\ell$ the angular momentum) and the Legendre polynomials. Therefore it permits 
to separate those features of the physical process that depend on the geometry 
or symmetry properties of the system, here represented by the Legendre 
polynomials, from those depending on the forces acting between the interacting
particles, here described by the partial--wave amplitudes. We are thus led to 
a separation between dynamics and symmetry. Accordingly, the classical nuclear 
theory describes the cross--section peak due to a resonance by fixed poles of 
the scattering amplitude in the complex momentum plane, which arise in complex 
conjugate pairs corresponding to $k$ and $-k^*$ ($k^2=E$ in suitable units), 
while the angular distribution is described by the Legendre polynomials. 
The theory does not attempt to group resonances in families, and provides only 
a local description in the neighbourhood of the energy position of the 
resonance. On the other hand, the phenomenology shows clearly that the 
resonances appear in ordered sequences, like rotational bands, which reflect 
the dynamical symmetries.

From the factorization between dynamics and symmetry it also follows that the 
angular distribution of the resonances, described by the Legendre polynomials 
(which are related to the unitary irreducible representation of the rotation 
group), cannot be in any way connected to the lifetime of the resonances. 
On the contrary, when the colliding particles are not identical, and therefore 
the scattering amplitude is not symmetrized (or antisymmetrized), the angular 
distribution retains memory of the incident beam direction and, consequently, 
it presents an asymmetry which is inversely related to the lifetime.

Another peculiar feature of the classical theory is that while the resonances 
are described by singularities of the scattering amplitude (Breit--Wigner poles), 
the antiresonances are described by the so--called \textit{potential scattering}, 
whose amplitude is the same of the one for the scattering by an impenetrable 
sphere: hard--core scattering. Resonances and antiresonances are depicted with 
completely different models. Furthermore, the exchange forces, which play a 
fundamental role in producing antiresonances, are not faithfully represented 
by the hard--core model. It follows that in the Breit--Wigner formalism 
the time delay due to the resonance cannot be neatly separated from the time 
advance due to the antiresonance.

Finally, the classical theory, which makes use of fixed poles, does not 
describe the dynamical evolution of resonances and antiresonances. Instead, 
the phenomenology shows that the resonance widths increase with energy, and 
the antiresonances tend to disappear at those energies where there is a smooth 
transition from quantum to semi--classical behaviour.

In this paper we try to recover the global character of the sequences of 
resonances, and specifically of the rotational bands. With this in mind, we 
shall represent the cross--section peaks due to resonances by the use of 
singularities of the scattering amplitude in the complex plane of the angular 
momentum, which is the generator of the rotation group. This can be achieved 
through a Watson--type resummation of the partial--wave expansion. Furthermore, 
we aim to describe resonances and antiresonances with analogous mathematical 
structures: i.e., scattering amplitude singularities in the complex angular 
momentum plane. 

In a previous paper \cite{DeMicheli2} -- hereafter referred as I -- we have 
studied ion collisions treating the composed structure of the clusters by means 
of the Jacobi coordinates and of the related $SU(n)$--group algebra. We proved 
that rotational bands emerge by removing the $SU(n)$ degeneracies by introducing 
forces that depend on the relative angular momentum of the clusters: i.e., 
non--local potentials. The latter is indeed generated by the exchange forces 
which enter the game precisely in connection with antiresonances. In the present 
paper we start from the results obtained in I; next we proceed to a Watson--type 
resummation of the partial--wave expansion. This procedure requires some 
additional constraints which limit the class of non--local potentials admitted.
We can thus study the singularities of the scattering amplitude in the complex 
angular momentum plane. We find that there exist two types of 
pole--singularities with $\Imag\lambda\neq 0$ ($\lambda$ denoting the complex 
angular momentum): the poles lying in the first quadrant (i.e., 
$\Imag\lambda > 0$), which describe the resonances; the poles lying in the 
fourth quadrant (i.e., $\Imag\lambda<0$), which correspond to antiresonances. 
These last poles are a peculiar property of non--local potentials. Another 
characteristic feature of the complex angular momentum poles is that their 
location varies as a function of the energy; they are \textit{moving poles} and, 
accordingly, we can speak of \textit{pole trajectories}. The main results which 
we prove in this paper are the following:
\begin{itemize}
\item[i)] Each rotational band of resonances is described by the trajectory 
of a single pole located in the first quadrant of the complex angular momentum 
plane (i.e., $\Imag\lambda>0$).
\item[ii)] The corresponding antiresonances are described by the trajectory 
of a single pole located in the fourth quadrant of the complex angular momentum 
plane (i.e., $\Imag\lambda<0$).
\item[iii)] The \textit{pure resonance} widths $\Gamma_R$ can be determined from 
the locations of poles in the first quadrant (resonance poles) and from their 
dependence on the energy.
\item[iv)] Both resonances and antiresonances are represented by pole 
singularities of the scattering amplitude: we thus obtain a unified scheme 
for describing time delay and time advance associated with the two processes.
\item[v)] We can describe the evolution of the rotational resonances into 
surface waves.
\end{itemize}
These results represent a very relevant improvement with respect to our previous 
analysis (see \cite{Fioravanti,Viano1,Viano2}). In fact, in these latter works 
the rotational resonances were fitted by using trajectories of the scattering 
amplitude poles lying in the first quadrant of the complex angular momentum 
plane, whereas the antiresonances were described by the hard--sphere potential 
scattering in a form very similar to the classical Breit--Wigner theory.

At this point we must strongly remark that $\Gamma_R$ cannot be identified with 
the width $\Gamma$ of the peak of the observed (experimental) cross--section. 
Indeed, the effect of the antiresonance distorts the bell--shaped structure of 
the pure resonance peak and, while we have a theoretical evaluation of 
$\Gamma_R$, we can only find an estimate of $\Gamma$ through statistical 
procedures, as it will be discussed in section \ref{se:phenomenological}.

In the complex angular momentum representation of the physical spectrum the 
bound states are described by pole singularities lying on the real positive 
semi--axis of the $\lambda$--plane (i.e., $\Real\lambda\geqslant 0,\,
\Imag\lambda=0$). Moving from bound states to resonances, the poles enter the 
first quadrant, $\Imag\lambda$ increases with the energy and describes the 
increase of the resonance widths $\Gamma_R$. At higher energy inelastic and 
reaction channels open, and the scattering target appears as a ball partially 
or totally opaque at the center: the resonances evolve into diffractive surface 
waves. This evolution is still described by the increase of $\Imag\lambda$ with 
energy. Similarly, the singularities in the fourth quadrant move away from the 
real axis, and at high energy their contribution becomes irrelevant. This
behaviour concords with the fact that the antiresonances, which are produced 
by exchange forces, are a quantum effect which disappears at the classical 
level.

The paper is organized as follows. In section \ref{se:outline} we outline the 
scattering theory for non--local potentials, summarizing the results obtained 
previously in I. In section \ref{se:complex} the singularities of the scattering 
amplitude in the complex angular momentum plane are studied; the formulae which 
represent the phase--shifts and that allow us to interpolate the rotational 
resonances, are given. In section \ref{se:from} we analyze the main properties 
of these singularities and their behaviour for high values of the energy; the 
transition from quantum--mechanical effects to classical behaviour is discussed 
and, accordingly, the evolution of rotational resonances in surface waves is 
described in detail. In section \ref{se:phenomenological} we present a 
phenomenological analysis in order to check the theory. In this section we 
reconsider the previous phenomenological work on the $\alpha$--$\alpha$ elastic 
collision \cite{Viano1}, and on the resonances and surface waves present in 
the $\pi^+$--p elastic scattering \cite{Viano2}.

\section{Outline of the scattering theory for non--local potentials}
\label{se:outline}  
In this section we sketch the main results obtained in I, which are essential 
to perform a Watson--type resummation of the partial--wave expansion, and that 
enable the analysis of the scattering amplitude singularities in the complex 
angular momentum plane.

In I we derived and studied the following integro--differential equation of 
Schr\"{o}dinger type:
\beq
(-\Delta +V_D)\chi(\bR)+g\int_{\R^3} V(\bR,\bR')\chi(\bR')\,\rmd\bR'= 
E \chi(\bR),
\label{1}
\eeq
which describes the interaction of two clusters. In (\ref{1}) $\Delta$ is the 
relative--motion kinetic energy operator, $g$ is a real coupling constant, 
$V_D$ is the potential which derives from direct forces, $V(\bR,\bR')$ 
represents the non--local potential which takes into account the exchange forces 
and, finally, $E$ represents, in the case of the scattering process, the 
relative kinetic energy of the two clusters in the center of mass system 
($\hbar=2\mu=1$, $\mu$ being the reduces mass of the clusters).
From the current conservation law it follows that $V(\bR,\bR')$ is a real and 
symmetric function: $V(\bR,\bR')=V^*(\bR,\bR')=V(\bR',\bR)$; moreover, 
$V(\bR,\bR')$ depends only on the lengths of the vectors $\bR,\bR'$ and on the 
angle $\gamma$ between them, or equivalently, on the dimension of the triangle
with vertices $(0,\bR,\bR')$ but not on its orientation. Hence, $V(\bR,\bR')$ 
can be formally expanded as follows:
\beq
V(\bR,\bR')=\frac{1}{4\pi R R'}\sum_{s=0}^\infty (2s+1) V_s(R,R') P_s(\cos\gamma),
\label{2}
\eeq
where $\cos\gamma=(\bR\cdot\bR')/(R R')$, $R=|\bR|$, and $P_s$ are the Legendre 
polynomials. The Fourier--Legendre coefficients $V_s(R,R')$ are given by:
\beq
V_s(R,R')=4\pi R R' \int_{-1}^1 V(R,R'; \cos\gamma) P_s(\cos\gamma)\, 
\rmd(\cos\gamma).
\label{3}
\eeq
We may therefore state that the l.h.s operator of equation (\ref{1}), acting on 
the wavefunction $\chi$, is a formally hermitian and rotationally invariant 
operator.

Next, we expand the relative--motion wavefunction $\chi(\bR)$ in the form:
\beq
\chi(\bR) = \frac{1}{R} \sum_{\ell=0}^\infty \chi_\ell(R) \, P_\ell(\cos\theta),
\label{4}
\eeq
where now $\ell$ is the relative angular momentum between the clusters.

Since $\gamma$ is the angle between the two vectors $\bR$ and $\bR'$, whose 
directions are determined by the angles $(\theta,\phi)$ and $(\theta',\phi')$ 
respectively, we have:
$\cos\gamma=\cos\theta\cos\theta' + \sin\theta\sin\theta' \cos(\phi-\phi')$. 
Then, by using the following addition formula for the Legendre polynomials:
\beq
\int_0^\pi \int_0^{2\pi} P_s(\cos\gamma)P_\ell(\cos\theta')\sin\theta'\, 
\rmd\theta'\, \rmd\phi'
= \frac{4\pi}{(2\ell+1)}P_\ell(\cos\theta)\delta_{s\ell}\,,
\label{5}
\eeq
from (\ref{1}), (\ref{2}), (\ref{4}), (\ref{5}) we obtain:
\beq
\chi''_\ell(R)+k^2 \chi_\ell(R) -\frac{\ell(\ell+1)}{R^2} \chi_\ell(R) = 
g\int_0^{+\infty} V_\ell(R,R') \chi_\ell(R') \, \rmd R',
\label{6}
\eeq
where $k^2=E$, and the local potential is now supposed to be included in the 
non--local one.

As in I, we suppose that $V(\bR,\bR')$ is a measurable function in 
$\R^3 \times \R^3$, and that there exists a constant $\alpha$ such that
\beq
C = \left\{\int_{\R^3} (1+R^2)e^{2\alpha R}\,\rmd\bR \int_{\R^3}(1+R'^2) R'^2 
e^{2\alpha R'}\, |V(\bR,\bR')|^2 \, \rmd\bR' \right\}^{1/2} < \infty.
\label{7}
\eeq
If bound (\ref{7}) is satisfied, then expansion (\ref{2}) converges in the norm 
$L^2(-1,1)$ for almost every $R$, $R' \in [0,+\infty)$ (the constant $\alpha$ 
will be used in the bound (\ref{14}) below). 

\begin{figure}[ht]
\centering
\includegraphics[scale=0.7]{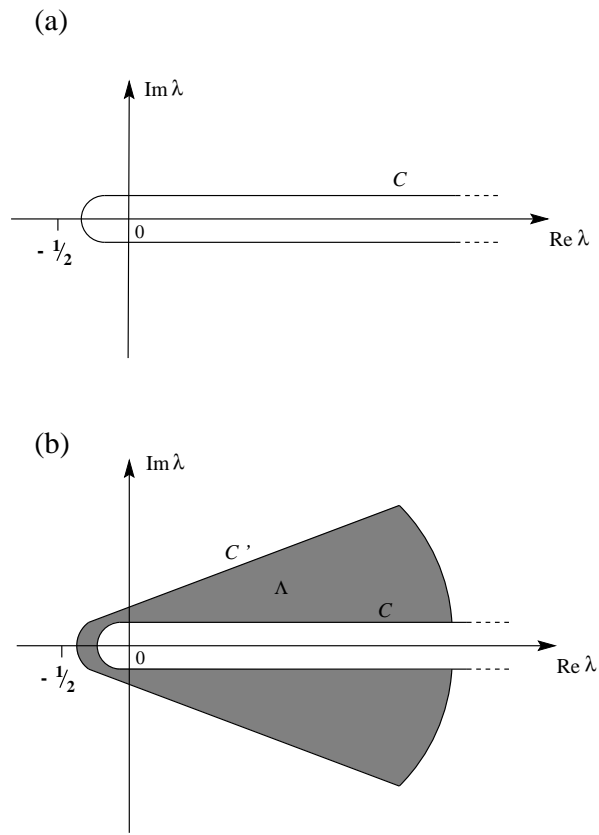}
\vspace*{7pt}
\caption{
\label{fig_1}
\small
(a) Integration path of integral in formula (\protect\ref{13}).
(b) Integration path of integral in formula (\protect\ref{15}).
}
\end{figure}

We must distinguish between two kinds of solutions to equation (\ref{6}): 
the scattering solutions $\chi_\ell^{(s)}(k,R)$, and the \textit{bound--state} 
solutions $\chi_\ell^{(b)}(R)$:
\begin{itemize}
\item[i)] The scattering solutions satisfy the following conditions:
\begin{subequations}
\label{8}
\begin{eqnarray}
&&\chi_\ell^{(s)}(k,R)=kRj_\ell(kR)+\Phi_\ell(k,R), \label{8a} \\
&&\sds\Phi_\ell(k,0)=0;~~\lim_{R\rightarrow +\infty}\left\{\frac{d}{dR}
\Phi_\ell(k,R)-\rmi k\Phi_\ell(k,R)\right\}=0, \label{8b}
\end{eqnarray}
\end{subequations}
where $j_\ell(kR)$ are the spherical Bessel functions, and the functions 
$d\Phi_\ell/dR$ are supposed to be absolutely continuous.
\item[ii)]
The \textit{bound--state} solutions $\chi_\ell^{(b)}(R)$ satisfy the conditions:
\beq
\int_0^{+\infty} \left | \chi_\ell^{(b)}(R) \right |^2 \, \rmd R < \infty,
~~~~~\chi_\ell^{(b)}(0)=0.
\label{9}
\eeq
\end{itemize}
Then, one can compare, as in the case of local potentials, the asymptotic 
behaviour of the scattering solution, for large values of $R$, with the 
asymptotic behaviour of the free radial function $j_\ell(kR)$, and 
correspondingly, introduce the phase--shifts $\delta_\ell(k)$; accordingly, 
we can define the scattering amplitude
\beq
T_\ell(k) = e^{\rmi\delta_\ell(k)}\,\sin\delta_\ell(k).
\label{10}
\eeq
Now, we can introduce, as in the standard collision theory, the total scattering 
amplitude that, in view of the rotational invariance of the total Hamiltonian, 
can be formally expanded in terms of Legendre polynomials as follows:
\beq
f(E,\theta) = \sum_{\ell=0}^\infty (2\ell+1) a_\ell(E) P_\ell(\cos\theta),
\label{11}
\eeq
where $E$ is the center of mass energy, $\theta$ is the center of mass scattering 
angle, $\ell$ is the relative angular momentum of the colliding clusters, and the 
partial scattering amplitudes $a_\ell(E)$ are given by:
\beq
a_\ell(E) = \frac{e^{2\rmi\delta_\ell}-1}{2\rmi k} = \frac{T_\ell(k)}{k}~,
~~~~~(k^2=E;\, 2\mu=\hbar=1).
\label{12}
\eeq
Expansion (\ref{11}) factorizes the amplitude into kinematics and dynamics: 
the Legendre polynomials $P_\ell(\cos\theta)$ (which are related to the unitary 
irreducible representation of the rotation group) describe the kinematics: 
the coefficients $a_\ell(E)$ (and specifically their singularities) reflect 
the dynamics.  On this factorization it is indeed based the classical 
Breit--Wigner theory of resonances, which separates kinematics from dynamics. 
But, as already mentioned, in this way the global aspect of the families of 
states, like the rotational bands, is lost. To recover the global aspect of 
the rotational bands, as suggested by the phenomenological data, one could try 
to represent the dynamics through the singularities (poles) in the complex plane
of the angular momentum, which is the generator of the rotation group. This can 
be achieved through a Watson resummation of expansion (\ref{11}), which 
consists in regarding the partial--wave series as an infinite sum over the 
residues of the poles of the function $1/\sin\pi\lambda$, obtained by the 
following integral whose integration path $C$ is shown in Fig. \ref{fig_1}a:
\beq
f(E,\theta)=\frac{\rmi}{2}\int_C\frac{(2\lambda +1)a(\lambda,E)
P_\lambda(-\cos\theta)}{\sin\pi\lambda}\,\rmd\lambda.
\label{13}
\eeq

\setlength{\unitlength}{1cm}

In order to deform conveniently the path $C$, it is usually assumed that the 
partial--waves $a_\ell(E)$ are the restriction to the integers of a function
$a(\lambda,E)$ ($\lambda\in\C,\,E$ fixed) meromorphic in the half--plane 
$\Real\lambda>-1/2$, holomorphic for $\Real\lambda>L-1/2$ (where $L$ is an 
integer larger than zero). All these properties are satisfied by the 
partial--waves associated with the class of the Yukawian potentials 
\cite{DeAlfaro}. In the case of non--local potentials some peculiar
features emerge because not only the centrifugal barrier, but also the potential 
itself depends on the angular momentum. Therefore, some results concerning the 
restriction on the position of the poles of the $S$--function 
$S(\lambda,k)=\exp [2\rmi\delta(\lambda,k)]$ $(\lambda\in\C,\,\Real\lambda >-1/2,
\, k$ real and fixed) fail in this case. We can note, as a typical 
example of this situation, that while for the class of the Yukawian potentials 
no poles of the $S$--function occur for $\Imag\lambda < 0$, this is not
the case for potentials which depend on the angular momentum.

By adding suitable constraints to the conditions introduced previously, one can 
prove that in the complex angular momentum plane there exists an angular sector 
$\Lambda$ where the partial amplitudes $a(\lambda,k)$ satisfy the 
following bound\footnote{This result, as well 
as the analytic continuation (Carlsonian interpolation of the 
\textit{partial potentials} $V_s(R,R')$ (see (\ref{3})), requires a rather delicate 
and detailed mathematical analysis, which we plan to discuss in a mathematical 
physics journal.}:
\beq
a(\lambda,k) = O(e^{-\beta\lambda}),~~~~~ 
\left(\lambda\in\Lambda,\,\Real\lambda\to +\infty;\,\cosh\beta=
1+\frac{2\alpha^2}{k^2}\right). 
\label{14}
\eeq
Accordingly, the contour $C$ can be deformed into the contour $C'$, as shown in 
Fig. \ref{fig_1}b. We thus obtain:
\beq
f(E,\theta)=\frac{\rmi}{2}\int_{C'}\frac{(2\lambda +1)a(\lambda,E)
P_\lambda(-\cos\theta)}{\sin\pi\lambda} \,\rmd\lambda
+\sum_{n=1}^N\frac{g_n(E)P_{\lambda_n}(-\cos\theta)}{\sin\pi\lambda_n(E)},
\label{15}
\eeq
where $\lambda_n(E) = \alpha_n(E)+\rmi\beta_n(E)$ give the location of the 
amplitude poles belonging to the angular sector $\Lambda$ of the $\lambda$--plane 
(see Fig. \ref{fig_1}b), and lying either in the first or in the fourth quadrant; 
$g_n(E)$ are the residues of $(2\ell+1)a_\ell(E)$ at the poles.
It is worth remarking that in the applications to low--energy nuclear physics, 
which are our concern, the path $C'$ not necessarily has to run parallel to the 
imaginary axis, as in the case of the high--energy physics. In fact, we are 
working in the physical region of $\cos\theta$ 
$(-1\leqslant\cos\theta\leqslant 1)$, and we are not interested to the 
asymptotic behaviour of the scattering amplitude for large transmitted momentum.

\section{Complex Angular Momentum Representation of Resonances and Antiresonances}
\label{se:complex}  
First let us consider the poles lying in the first quadrant and within the 
angular sector $\Lambda$ (see Fig. \ref{fig_1}b); suppose that at a certain energy, and 
for a specific value $n_0$ of $n$, $\alpha_{n_0}$ crosses an integer, while 
$\beta_{n_0} \ll 1$; then the corresponding term in the sum over the poles in 
representation (\ref{15}) becomes very large: we have a pole dominance.
Therefore, in the neighbourhood of a sharp and isolated resonance the following 
approximation for the amplitude is worth trying:
\beq
f(E,\theta) \simeq g(E) \frac{P_\lambda(-\cos\theta)}{\sin\pi\lambda(E)},
\label{16}
\eeq
(for simplicity, we have dropped the subscript $n_0$).
The Legendre function $P_\lambda(-\cos\theta)$ presents a logarithmic singularity 
at $\theta = 0$ \cite{Bateman,Sommerfeld}. Therefore, approximation 
(\ref{16}) certainly breaks down forwards, where it is necessary to take 
into account also the contribution of the background integral, in order to make 
the amplitude $f(E,\theta)$ finite and regular. Conversely, approximation 
(\ref{16}) is satisfactory at backward angles. The logarithmic singularity, 
that the Legendre functions $P_\lambda(-\cos\theta)$ $(\lambda\in\C)$ present at 
$\theta = 0$, clearly indicates that the angular distribution given by these 
functions, and, accordingly, by approximation (\ref{16}), is asymmetric.
However, let us observe that this peculiar feature of approximation 
(\ref{16}) should not be regarded as a defect, but it could be interpreted 
as the capability of this representation of displaying the specific
aspect of the resonances of non--identical particles, i.e. the asymmetry proper 
of the unstable states. Since the angular lifetime is finite, the isotropy in 
$\theta$ is broken: there is memory of the incident beam direction. 
The angular asymmetry proper of the resonances can then be associated with the 
\textit{spin--width} of the compound state (see I, section 4): to small values of 
angular momentum dispersion there corresponds a small degree of asymmetry in the 
angular distribution.

Amplitude (\ref{16}) diverges logarithmically forwards and the differential 
cross--section diverges as the square of the logarithm, but the total 
cross--section, derived from (\ref{16}), is finite. Indeed, we have:
\beq
\sigma_{tot.}=\frac{2\pi|g(E)|^2}{|\sin\pi\lambda(E)|^2}\,
\int_0^\pi |P_\lambda(-\cos\theta)|^2\sin\theta\,\rmd\theta,
\label{17}
\eeq
and the integral at the r.h.s. of formula (\ref{17}) converges. Furthermore, 
we may project the amplitude (\ref{16}) on the $\ell$--th partial--wave, 
obtaining:
\beq
a_\ell=\frac{e^{2\rmi\delta_\ell}-1}{2\rmi k}=
\frac{g}{\pi}\frac{1}{(\alpha_R+\rmi\beta_R-\ell)(\alpha_R+\rmi\beta_R+\ell+1)},
\label{18}
\eeq
where we write $\lambda=\alpha_R+\rmi\beta_R$ to emphasize that we are now 
referring to resonances. Next, when the elastic unitarity condition can be 
applied, we have the following relationship among $g$, $\alpha_R$ and $\beta_R$:
\beq
g = -\frac{\pi}{k}\beta_R (2\alpha_R+1),
\label{19}
\eeq
and, finally, we obtain:
\beq
\delta_\ell=\sin^{-1}\frac{\beta_R(2\alpha_R+1)}{\{[(\ell-\alpha_R)^2+\beta_R^2]
[(\ell+\alpha_R+1)^2+\beta_R^2]\}^{1/2}}.
\label{20}
\eeq

\smallspace

\underline{Remark}: Let us note that the phase--shifts of formula (\ref{20}) 
do not satisfy the asymptotic behaviour, for large values of $\ell$, prescribed 
by formula (\ref{14}). This derives from the fact that the amplitude 
given by formula (\ref{16}), and, accordingly, the phase--shifts
given by formula (\ref{20}), are approximations which are faithful only for 
small values of $\ell$. These approximations, however, are acceptable at low 
energy where a few terms of the partial--wave expansion are sufficient to 
describe the scattering amplitude.

\smallspace

Representation (\ref{20}) is useful for the following reasons:
\begin{itemize}
\item[i)] at fixed energy it gives several phase--shifts at different values of 
$\ell$, with an acceptable approximation for small values of $\ell$.
\item[ii)] $\alpha_R$ and $\beta_R$ depend on the energy $E$: i.e., we have a 
pole trajectory; when $\alpha_R(E)$ equals an integer $\ell$, and $\beta_R(E)$ 
is very small, we have $\sin\delta_\ell \simeq 1$, i.e. a resonance.
\item[iii)] Representation (\ref{20}) can describe, in principle, a sequence 
of resonances in various partial--waves.
\end{itemize}

Next, if we assume that the colliding particle are spinless bosons (e.g., 
$\alpha$ particles in $\alpha$--$\alpha$ collision), then the scattering 
amplitude must be symmetrized. Therefore, instead of approximation 
(\ref{16}), we must write the following one:
\beq
f(E,\theta)\simeq \frac{g(E)}{\sin\pi\lambda(E)} \left[
\frac{P_\lambda(\cos\theta)+P_\lambda(-\cos\theta)}{2}\right],
\label{21}
\eeq
and, consequently:
\beq
\delta_\ell=\sin^{-1}\left\{\frac{1+(-1)^\ell}{2}
\frac{\beta_R(2\alpha_R+1)}{\{[(\ell-\alpha_R)^2+\beta_R^2]
[(\ell+\alpha_R+1)^2+\beta_R^2]\}^{1/2}} \right\}.
\label{22}
\eeq
Approximation (\ref{21}) fails at $\theta=0$ and $\theta=\pi$, and, in view 
of the combination of $P_\lambda(\cos\theta)$ and $P_\lambda(-\cos\theta)$, the 
angular pattern showed by approximation (\ref{21}) is symmetric.

Coming back to formula (\ref{18}), and expanding the term 
$\lambda(E)=\alpha_R(E)+\rmi\beta_R(E)$ in Taylor's series in the neighbourhood 
of the resonance energy, we can derive an estimate of the resonance width 
$\Gamma_R$: i.e.,
\beq
\Gamma_R=\frac{2\beta_R(\textrm{d}\alpha_R/\textrm{d}E)}
{(\textrm{d}\alpha_R/\textrm{d}E)^2+ 
(\textrm{d}\beta_R/\textrm{d}E)^2}.
\label{23}
\eeq
However, $\Gamma_R$ should not be identified with the width $\Gamma$ of the 
observed peak of the cross-section (see below at the end of this section, 
and the next section \ref{se:phenomenological}). We can now give the following 
physical interpretation of the term $\beta_R$. The resonance is unstable for the 
leakage of particles tunnelling across the centrifugal barrier. Therefore the 
probability of finding the particles inside a given sphere must decrease with 
time. The only possibility of keeping up with the loss of probability is to 
introduce a source \textit{somewhere}. The source can be provided by the complex 
centrifugal barrier. The condition for the source to be emitting is just to take 
$\Imag\lambda\equiv\beta_R$ positive (see also Eq. (\ref{24}) below).

We now focus on a peculiar feature associated with non--local potentials:
in addition to poles located in the first quadrant (i.e., $\Imag\lambda > 0$), 
we have also poles located in the fourth quadrant (i.e., $\Imag\lambda < 0$), 
which describe the antiresonances. In order to achieve a better understanding
of this point, let us consider the continuity equation, which reads (see also 
equation (95) of I):
\beq
\frac{\partial w}{\partial t} + \nabla\cdot\bj = 2 w \,\Imag {\cU}_\textrm{eff},
\label{24}
\eeq
where $w=\chi^* \chi$, $\chi$ being the wavefunction (see also (\ref{4})), 
$\bj$ is the current density: i.e., 
$\bj=\rmi\{\chi\nabla\chi^*-\chi^*\nabla\chi\}$, and $\cU_\textrm{eff}$ is the sum 
of the potential and of the centrifugal term. In the case of local potentials, 
which do not depend on the angular momentum, the contribution to 
$\Imag {\cU}_\textrm{eff}$ comes only from the extension of the centrifugal term to 
complex--valued angular momentum. The resonances are precisely related to this 
term. When the potential is non--local, and therefore depends on the angular 
momentum, a contribution to $\Imag {\cU}_\textrm{eff}$ comes also from the 
potential when the angular momentum is extended to complex values. This 
contribution cannot be in any way related to the centrifugal barrier, and 
therefore it cannot be connected to resonances. It rather derives from 
non--locality and, accordingly, from the presence of exchange forces which 
generates the antiresonances (see I). In this case the probability of finding 
the scattered particle inside a given sphere increases with time. The only 
possibility of keeping up with this increase of probability is to introduce a 
negative source: i.e., a pole singularity with $\Imag\lambda<0$. This term 
cannot in any way be connected to a (metastable) state; however, we still have 
a dispersion in energy according to the uncertainty principle. We speak in this 
case of \textit{time advance} instead of \textit{time delay}, which is proper of the 
resonance. Analogously, we have a connection between dispersion in angle and 
dispersion in angular momentum. In fact, let us remark that the antiresonances 
are a typical quantum--mechanical effect due to the exchange forces.
Then, proceeding exactly as in the case of resonances, and denoting the poles in 
the fourth quadrant by $\lambda=\alpha_A-\rmi\beta_A$, $\beta_A >0$, we have in 
the neighbourhood of an antiresonance:
\beq
\delta_\ell=\sin^{-1}
\frac{-\beta_A(2\alpha_A+1)}{\{[(\ell-\alpha_A)^2+\beta_A^2]
[(\ell+\alpha_A+1)^2+\beta_A^2]\}^{1/2}},
\label{25}
\eeq
or, in the case of identical scalar particles:
\beq
\delta_\ell=\sin^{-1}\left\{\frac{1+(-1)^\ell}{2}
\frac{-\beta_A(2\alpha_A+1)}{\{[(\ell-\alpha_A)^2+\beta_A^2][(\ell+\alpha_A+1)^2
+\beta_A^2]\}^{1/2}}\right\}.
\label{26}
\eeq
Adding the contributions of poles lying in the first and fourth quadrant, 
we have:
\begin{eqnarray}
\label{27}
\delta_\ell &=& \sin^{-1}\left\{
\frac{\beta_R (2\alpha_R+1)}
{\{[(\ell-\alpha_R)^2+\beta_R^2][(\ell+\alpha_R+1)^2+\beta_R^2]\}^{1/2}}\right\} 
\nonumber \\
&+&\sin^{-1}\left\{\frac{-\beta_A(2\alpha_A+1)}
{\{[(\ell-\alpha_A)^2+\beta_A^2][(\ell+\alpha_A+1)^2+\beta_A^2]\}^{1/2}}\right\},
\end{eqnarray}
and, in the case of identical scalar particles we obtain:
\begin{eqnarray}
\label{28}
\delta_\ell&=&\sin^{-1}\left\{\frac{1+(-1)^\ell}{2}
\frac{\beta_R (2\alpha_R+1)}{\{[(\ell-\alpha_R)^2+\beta_R^2]
[(\ell+\alpha_R+1)^2+\beta_R^2]\}^{1/2}}\right\} \nonumber \\
&+&\sin^{-1}\left\{\frac{1+(-1)^\ell}{2}
\frac{-\beta_A(2\alpha_A+1)}{\{[(\ell-\alpha_A)^2+\beta_A^2]
[(\ell+\alpha_A+1)^2+\beta_A^2]\}^{1/2}}\right\}.
\end{eqnarray}

Finally, we recall the expression of the total cross--section in terms of 
phase--shifts:
\beq
\sigma_\textrm{tot} = \frac{4\pi}{k^2} \sum_{\ell=0}^\infty (2\ell+1) 
\sin^2\delta_\ell.
\label{29}
\eeq

In principle one could try to introduce an estimate of the width of the 
antiresonances in a form similar to that followed in the case of resonances. 
In other words we could expand in Taylor's series the term 
$\lambda(E) = \alpha_A(E)-\rmi\beta_A(E)$, in the neighbourhood of the 
antiresonance energy. These widths should be inversely proportional to the time 
advance of the outgoing flux. But, while the resonances produce sharp peaks in 
the cross--section and the width of the peaks is well defined and could be 
properly estimated, this is not the case for the antiresonances. The 
contribution of the antiresonances to the cross--section does not produce sharp 
peaks and, correspondingly, the width of the antiresonances is ill--defined; 
instead, the antiresonances are responsible of the asymmetry of the 
cross--section peaks (see section \ref{se:phenomenological} for phenomenological 
analysis and examples).
Moreover, we want strongly remark that in the present procedure we can separate 
neatly the contribution of the resonance from that of the antiresonance and, 
accordingly, the time delay from the time advance. Resonances and antiresonances 
are both described in terms of pole singularities, \textit{but acting at different 
values of energy}. It follows that the interference effect between these terms
appears negligible as it will be shown phenomenologically in section 
\ref{se:phenomenological}. In view of the difficulty of defining an 
\textit{antiresonance width} $\Gamma_A$, it remains the problem of finding an 
estimate of the \textit{total} width $\Gamma$ of the observed (experimental)
resonance peak, which is larger than $\Gamma_R$ in view of the distorting effect 
of the antiresonance. This problem will be analyzed and discussed in connection 
with the phenomenological analysis in section \ref{se:phenomenological}.

\section{From Resonances to Surface Waves}
\label{se:from}  
Let us firstly consider the case of Yukawian local potentials, which are 
represented by a spherically symmetric function on $\R^3$ of the following form: 
$V(r)=\frac{1}{r}\int_{\mu_0}^{+\infty}e^{-\mu r}\sigma(\mu)\,\rmd\mu$, 
($\mu_0>0$). We then recall the main properties of the pole trajectories 
generated by this class of potentials. At negative values of the energy $E=k^2$, 
the poles $\lambda_n(k)$ lie on the real axis of the $\lambda$--plane, and 
describe bound states. The bound state wavefunction belongs to $L^2(0,+\infty)$.
Further, these states are stable and, accordingly, they are described either by 
poles lying on the imaginary axis of the upper half of the $k$--plane
(i.e., $\Real k = 0$), or by poles located on the real axis of the 
$\lambda$--plane (i.e., $\Imag \lambda = 0$): their lifetime is infinite. 
Increasing the energy, the poles in the $\lambda$--plane move to the right. 
This is nothing but the familiar fact that the binding energies of the 
corresponding bound states of different angular momenta must decrease with 
increasing $\ell$. At positive energy $E = k^2$, the poles $\lambda_n(k)$ enter 
the first quadrant of the complex $\lambda$--plane: at those values where
$\Real\lambda_n(k)$ cross an integer $\ell$ and correspondingly 
$\Imag\lambda_n(k)$ is small, we have a resonance. The lifetime of the latter 
is inversely proportional to $\Imag\lambda_n(k)$. As the energy increases the
trajectory could in principle describe several resonances: each of them 
corresponding to the crossing of $\Real\lambda_n(k)$ through an integer value 
$\ell$. We could thus have several resonances lying on the same trajectory. 
But, as we shall show below, this is not the case. In fact, one can show 
numerical examples of trajectories produced by Yukawian potentials that leave 
very soon the real $\lambda$--axis, then turn back toward the left half of the 
$\lambda$--plane, and do not exhibit an interpolation of resonances 
\cite{Nussenzveig1}. Furthermore, it can be proved \cite{DeAlfaro,Nussenzveig1} 
that each trajectory, produced by the Yukawian potential, necessarily turns back 
toward the left half of the $\lambda$--plane. Therefore, if a pole--trajectory 
$\lambda_n(k)$, moving with positive derivative 
$\frac{d\,\Real\lambda_n(k)}{dk}$, goes through an integer value
$\ell$ (i.e., $\Real\lambda_n(k)=\ell$) and produces a resonance, then, after 
having turned back to the left, must necessarily pass through the same integer 
value $\ell$, but now with negative derivative. This latter crossing is not 
associated with a time delay, but with an advance of the outgoing wave. Thus 
the peak in the cross--section is not a resonance, but comes from the downward 
passage of the phase--shift through $\pi/2$: i.e., it corresponds to an 
antiresonance. Let us now suppose that a pole--trajectory $\lambda_n(k)$,
moving with positive derivative $\frac{d\Real\lambda_n(k)}{dk}$, crosses 
several integer values $\ell$ and describes a sequence of resonances, ordered 
according to the increasing values of the angular momentum. Then, the same 
trajectory, after having turned back to the left half of the $\lambda$--plane, 
should necessarily describe the corresponding antiresonances in inverse order. 
One would have at a smaller value of the energy the antiresonances which 
correspond to the resonances with higher angular momentum. \textit{To an ordered 
sequence of resonances (the order being given by the angular momentum) it would 
correspond a sequence of antiresonances ordered inversely}. This is manifestly 
contradictory. We can thus conclude that it is not possible to describe with a 
pole--trajectory $\lambda_n(k)$ the ordered sequences of resonances and 
antiresonances within the framework of the Yukawian potentials. 

Therefore, if we want to describe the ordered sequences of resonances and 
antiresonances, like those produced by rotational bands, we are forced to refer 
to a larger class of potentials, like the non--local ones, introduced previously. 
As we have seen, this class of potentials admit poles in the fourth quadrant, 
which are proper for describing antiresonances in view of the fact that their 
imaginary part is negative, and therefore can describe a time advance instead of 
a time delay. 

In conclusion, we have poles in the first quadrant of the $\lambda$--plane, 
whose trajectories describe ordered sequence of resonances, and poles in the 
fourth quadrant whose trajectories describe ordered sequence of antiresonances. 
The locations of the poles lying in the first quadrant have an imaginary part 
that increases with energy, and this behaviour corresponds to the increase of 
the widths of the rotational resonances. The trajectories of these poles keep 
moving to the right, and not necessarily should turn back to the left.
In this evolution we pass from a pure quantum--mechanical effect to 
semi--classical effects, which can be described in terms of surface waves, 
as we shall explain below.

As the energy increases, inelastic and reaction channels open: the scenario 
drastically changes. The elastic unitarity condition does not hold anymore, 
and the target may now be thought of as a ball partially or totally opaque at 
the center, and with a \textit{semitransparent} shell at the border. Accordingly, 
the potential acquires an imaginary part. The structure of the singularities 
becomes much more complicated than in the one--channel case.  In fact, the 
structure of the singularities is a superposition of cuts starting at every 
threshold $E_{\gamma_i}$, $\gamma_i$ labelling the channels. In spite of these 
difficulties, we can still, roughly speaking, continue the trajectory of the 
pole in the complex angular momentum plane with the only caution of avoiding the
use of the elastic unitarity condition (\ref{19}). Moreover, when the energy increases,
the effects of the exchange forces, which are a pure quantum effect, 
tend to vanish. Therefore we can still retain approximation (\ref{16}), 
which represents the elastic scattering, but now its physical interpretation must 
be appropriately modified. We have seen that the width of the rotational 
resonances increases with the energy; accordingly, $\beta_R\equiv\Imag\lambda$ 
increases. It follows that when $\alpha_R\equiv\Real\lambda$ crosses an integer 
value, but $\beta_R$ is not much smaller than one, we do not observe sharp peaks 
in the cross--section since 
$|\sin\pi(\alpha_R+\rmi\beta_R)|^{-1}\simeq\exp\{-\pi\beta_R\}$: we have 
diffractive effects. These phenomena can be very well described by returning to 
formula (\ref{16}), which can now be conveniently written in the following 
form:
\beq
f(E,\theta) \simeq C(E)\,P_{\lambda(E)}(-\cos\theta).
\label{30}
\eeq
In this way the amplitude is factorized into two terms: the first one, $C(E)$, 
gives the amplitude at $\theta=\pi$ as a function of $E$ (in fact, 
$P_\lambda(1)=1$); the second factor describes the backward angular distribution 
at fixed $E$. This second term may be easily interpreted by recalling the 
asymptotic behaviour of $P_{\lambda}(-\cos\theta)$ for large values of 
$|\lambda|$. In fact, it holds \cite{Bateman}:
\beq
P_\lambda(-\cos\theta)\propto
\frac{e^{-\rmi[(\lambda+1/2)(\pi-\theta)-\pi/4]}+
e^{\rmi[(\lambda+1/2)(\pi-\theta)-\pi/4]}}
{(2\pi\lambda\sin\theta)^{1/2}},~~~~~(0<\theta<\pi).
\label{31}
\eeq
These exponentials correspond to the surface waves excited at the periphery of the 
target by the grazing rays. These rays undergo at the point of tangency a 
splitting: one ray leaves the target (which is supposed, for simplicity, to
be a sphere) tangentially, while the other one propagates along the edge; 
finally, if the black--body limit is not yet reached, we may also have a 
refracted ray which penetrates the weakly absorbing region. Moreover, the 
surface ray undergoes at any point the same splitting described above; also the
refracted ray, after emerging at the surface, undergoes the same splitting, 
sending off tangential rays. However, these latter rays are in phase and 
unidirectional only in the direct forward and backward directions. The factor 
$(\sin\theta)^{-1/2}$ describes precisely this focusing effect at $\theta=0$, 
and $\theta=\pi$.

In order to have the exact expression of the surface waves, excited by the 
grazing rays at the periphery of the target, we must introduce the surface angle 
$\theta_m^{(S_\pm)}$ in place of the scattering angle $\theta$.
The relationship is as follows:
\begin{subequations}
\label{32}
\begin{eqnarray}
\theta_m^{(S_+)} &=& \theta + 2\pi m,~~~~~(m=0,1,2,\ldots), \label{32a} \\
\theta_m^{(S_-)} &=& 2\pi-\theta + 2\pi m, \label{32b}
\end{eqnarray}
\end{subequations}
where $\theta_m^{(S_+)}$ refers to the counterclockwise travelling rays, while 
$\theta_m^{(S_-)}$ corresponds to the clockwise ones. By using formula 
(\ref{31}) it can be shown that the Legendre function $P_\lambda(-\cos\theta)$ 
can be generated by the superposition of exponentials of the form 
$e^{\rmi\lambda\theta_m^{(S_\pm)}}$, representing surface waves creeping around 
the target \cite{DeMicheli1}. Coming back to formula (\ref{16}), 
we recall that the latter is logarithmically divergent at $\theta=0$, but it is 
not at $\theta=\pi$. This different behaviour is in agreement with the fact that 
the damping factor of the surface waves is given by: 
$e^{-\Imag\lambda\theta_m^{(S_\pm)}}$. Then, forwards, where the damping 
factor is nearly equal to 1, an infinity of creeping waves contribute to the 
total amplitude: we have a divergence. On the contrary, the creeping wave 
approximation works better backwards, since $\Imag\lambda\,\theta_m^{(S_\pm)}$
is larger.

\begin{figure}[ht]
\centering
\includegraphics[scale=0.7]{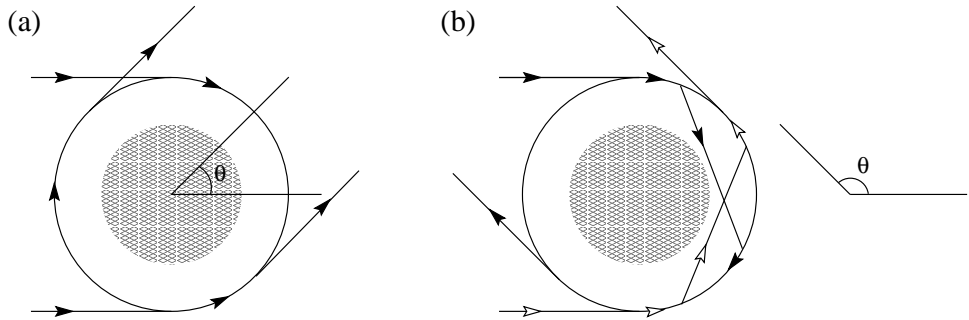}
\vspace*{7pt}
\caption{
\label{fig_2}
\small
Diffracted rays emerging along the direction $\theta$.
(a) Without taking any shortcut.
(b) Taking one shortcut.
}
\end{figure}

In order to evaluate the term $C(E)$, let us suppose that the black--body limit 
has been reached. Then the grazing ray at the tangency point will split into two 
branches only: a surface ray which describes a geodesic around the target, and 
another ray which leaves the surface tangentially. The point where diffraction
takes place may be regarded as an interaction vertex, characterized by a 
coupling constant, while the line joining two vertices may be regarded as a 
propagator. The coupling constants are the diffraction coefficients; the 
propagator, for a surface ray describing an arc length $\theta_m^{(S_\pm)}$, 
takes the form $e^{\rmi\lambda\theta_m^{(S_\pm)}}$. Next, we suppose that the 
so--called \textit{localization principle} holds true: the phenomena which occur 
for large energy, on the periphery of the interaction region, are independent 
of the inner structure of the target. Therefore we take for $\Imag\lambda$ 
(which characterizes the propagator), as well as for the coupling constant
(the diffraction coefficient) an energy dependence as the one calculated for a 
wholly transparent sphere \cite{Nussenzveig2}. Then, the formula 
for the cross--section at $\theta=\pi$ is given by \cite{Nussenzveig2,Viano2}:
\beq
\frac{\pi}{k^2}\left(\frac{\rmd\sigma}{\rmd\Omega}\right)_{\theta=\pi} = 
C_0\,k^{-4/3}\,e^{-c\,k^{1/3}}, ~~~~~(c = \Cons),
\label{33}
\eeq
where $C_0$ is a constant beyond the black--body limit. Before reaching the 
black--body limit, the grazing rays may undergo a critical refraction, penetrate 
the weakly absorbing region, and then emerge after one or two shortcuts 
(see Fig. \ref{fig_2}). In this case $C_0$ is not a constant since it must take into
account the contributions of various components: the diffracted rays which do 
not undergo any shortcut and the critically refracted rays which take one or two 
shortcuts before emerging. These various contributions interfere, producing an 
oscillating pattern. But, as the momentum increases, the radius of the central 
opaque core increases too, and the shortcuts are progressively suppressed; 
therefore the amplitude of these oscillations is damped and tends to vanish 
towards the black--body limit. In order to fit this oscillating pattern, we use 
a function of the following form: $(1+A\sin(\omega k+\phi))$. Accordingly, 
we shall fit the experimental data (see next section \ref{se:phenomenological}) 
with the following formula:
\beq
\frac{\pi}{k^2}\left(\frac{\rmd\sigma}{\rmd\Omega}\right)_{\theta=\pi} = 
C_0\,k^{-4/3}\,e^{-c\,k^{1/3}} (1+A\sin(\omega k+\phi)).
\label{34}
\eeq

\setlength{\unitlength}{1cm}

Let us finally note that the complex angular momentum poles which are connected 
to resonances and those which are associated with surface waves present 
remarkable differences. We can, indeed, speak of two different classes of poles 
\cite{Nussenzveig1}. The poles of the first class are located near the real axis
of the complex $\lambda$--plane, and are associated with resonances. The poles 
of the second class lie along a line which is nearly parallel to the imaginary 
axis: they are insensitive to the behaviour of the potential in the inner region, 
and are associated with surface waves \cite{Nussenzveig1}. Further, the poles 
of the second class move approximately parallel to the real axis when the energy
increases.

To these different classes of poles we can associate two different physical 
models: vortices and surface waves produced by diffracted rays. In order to 
summarize rapidly the hydrodynamical model of the vortices, we move back to the 
expression of the current density, introduced in section \ref{se:complex}: i.e., 
$\bj=\rmi\{\chi\nabla\chi^*-\chi^*\nabla\chi\}$, where $\chi$ is the 
wavefunction (see also Ref. \cite{DeMicheli2}). Assuming a semiclassical 
approximation, we write $\chi = \frac{A}{\sqrt{2}}e^{\rmi\Theta}$ ($A=$ constant);
accordingly, we have: $\bj=A^2\nabla\Theta$. Then we introduce a velocity field 
$\mathbf{v}$, regarding $\Theta$ as a velocity potential in the hypothesis of 
irrotational flow: $\mathbf{v}=\nabla\Theta$. First, we represent the incoming beam 
as an irrotational flow streaming around the target. Then the trapping proper 
of the resonance can be depicted as a rotational flow $\mathbf{\omega}$ given by: 
$\mathbf{\omega} = \nabla \times \mathbf{v}$.

The diffracted rays which generate surface waves are due to a completely 
different process. Regarding the diffraction as an obstacle problem in a 
Riemannian space with boundary, we can consider the edge of the diffracting body 
as the boundary of the ambient space. Then the determination of the geodesics by 
their initial tangent (Cauchy problem) is not unique: when a ray grazes a 
boundary surface, the ray splits in two parts, one keeps going as an ordinary 
ray, whereas the other part travels along the surface. This is precisely the 
mechanism which generates diffracted rays and surface waves \cite{DeMicheli1}. 
We thus call the first class of poles associated with vortices, and representing 
resonances, Regge poles; while the name Sommerfeld poles is deserved to the second 
class of poles, in view of the fact that they have been discovered by Sommerfeld 
\cite{Sommerfeld} in connection with the diffraction of radio waves around the 
earth.

\section{Phenomenological analysis}
\label{se:phenomenological}  
\subsection{Rotational band in \boldmath{$\alpha$}--\boldmath{$\alpha$} elastic scattering}
\label{subse:alphaalpha}
The $\alpha$--$\alpha$ elastic scattering is certainly a good laboratory for 
testing the theory, since it is a system of two identical spinless particles 
which clearly displays rotational bands. The data, which we analyze, are the 
experimental phase--shifts, taken from Refs. 
\cite{Afzal,Buck,Chien,Darriulat,Tombrello}, the range of energies extending up 
to 40 MeV in the center of mass system. In this range of energies, in addition 
to the elastic channel that we here consider (notice that to obtain formula 
(\ref{28}) the unitary elastic condition was assumed to hold), there are 
several other inelastic and reaction channels: for instance, the reaction 
channel p+Li$^7$, whose threshold energy is 17.25 MeV \cite{Okai}. 
Therefore, the experimental phase--shifts contain non--vanishing imaginary parts, 
and even their real part is affected by these channels. For this reason 
considering only the real part of the experimental phase--shifts, as we do in 
the following, is admittedly an approximation. Nevertheless, we assume that 
the effects of the non--elastic channels on the real part of the experimental 
phase--shifts is negligible within the considered energy range, and we compare 
the phase--shifts computed by the theory presented in the previous sections with 
the real part of the experimental phase--shifts. Finally, instead of fitting 
the differential cross--section, we prefer to fit the phase--shifts so that the 
action of the Coulomb potential can be subtracted.
However, in connection with the interference effects related to the Coulomb subtraction,
the following two remarks should be made: \\
i) In view of the long range of the Coulomb force, the exchange part of the
Coulomb interaction does not affect greatly the scattering wavefunction 
(see Ref. \cite{Wildermuth} and the references therein)\footnote{For 
a more detailed analysis of this topic and for a numerical comparison between the 
$\alpha$--$\alpha$ phase--shifts computed with and without the exact
exchange Coulomb interaction, the interested reader is referred to the 
appendix B of Ref. \cite{Tang} (see, in particular, Fig. 21 of this appendix).}. \\
ii) For the sake of preciseness, one should distinguish the 
\textit{quasi--nuclear} phase--shifts $\delta_\ell$ from the \textit{purely nuclear}
phase--shifts $\delta_\ell^*$, which are those associated with the scattering between 
the same particles, with the same strong properties but without the Coulomb
interaction \cite{Reignier1}. It is rather intuitive, and can be rigorously proved,
\cite{Reignier1,Reignier2} that they differ by quantities proportional to the
Sommerfeld parameter $\eta=\frac{Z_1Z_2e^2}{\hbar v}$. Even though $\eta$ can be
significantly large at low energies, nevertheless this fact does not prevent 
the phase--shifts $\delta_\ell$ from being treated as corresponding to a short range 
potential and retaining the main properties of interest for our analysis.
The reader interested to a rigorous mathematical analysis of the Coulomb effects
at $k=0$ is referred to Ref. \cite{Bertero}, where it is proved that an additional
Coulomb potential does not affect the general properties of the Regge's trajectories
except for their threshold behaviour.

The functions $\alpha_R(E)$, $\beta_R(E)$, $\alpha_A(E)$ and 
$\beta_A(E)$ in formula (\ref{28}) are parametrized as follows: 
\begin{subequations}
\label{35}
\begin{eqnarray}
&&\alpha_R(E)[\alpha_R(E)+1] = 2IE + \alpha_0, \label{35a} \\
&&\beta_R(E) = b_1 \sqrt{E}, \label{35b} \\
&&\alpha_A(E) =  a_1 E^{1/4}, \label{35c} \\
&&\beta_A(E) = g_0 (1-e^{-E/E_0}) + g_1 E + g_2 E^2, \label{35d}
\end{eqnarray}
\end{subequations}
where $I=\mu R^2$ is the moment of inertia, $\mu$ is the reduced mass, $R$ is 
the interparticle distance, and $E$ is the center of mass energy. Formula 
(\ref{35a}) for the resonant component $\alpha_R(E)$ simply gives the
angular momentum of the two--particle system viewed, in first approximation, 
as a rotator; formula (\ref{35b}) states for $\beta(E)$ a growth which is 
fast for low energy, but slower for higher energy; this behaviour suits the 
analysis done in section \ref{se:from} concerning the evolution of the resonances 
into surface waves for sufficiently large energy. For what concerns $\beta_A(E)$ 
in formula (\ref{35d}), the role of the exponential term is just to 
make a smooth, though rapid, transition of $\beta_A(E)$ from zero to the constant 
$g_0$ so to have a regular behaviour at very low energy. Unfortunately, a model 
which prescribes the growth properties of $\alpha_A(E)$ and $\beta_A(E)$ is, 
at present, missing. This would require a refined theory able to describe
the evolution toward semiclassical and classical phenomena.
The quantities $I$, $\alpha_0$, $b_1$, $a_1$, $g_0$, $g_1$ and $g_2$ should be 
regarded as fitting parameters. In what follows we will consider phenomena 
occurring only above threshold; therefore we do not analyze the $\ell=0$ 
phase--shift, which is ruled by the low--energy behavior of the trajectories 
$\alpha_A(E)$ and $\beta_A(E)$. In particular, the resonance--antiresonance 
correspondence is missing in the $\ell=0$ phase--shift, and to date this makes 
it difficult to reproduce this phase--shift in the framework of our model.

\setlength{\unitlength}{1cm}
\begin{figure}[th]
\centering
\includegraphics[scale=1.0]{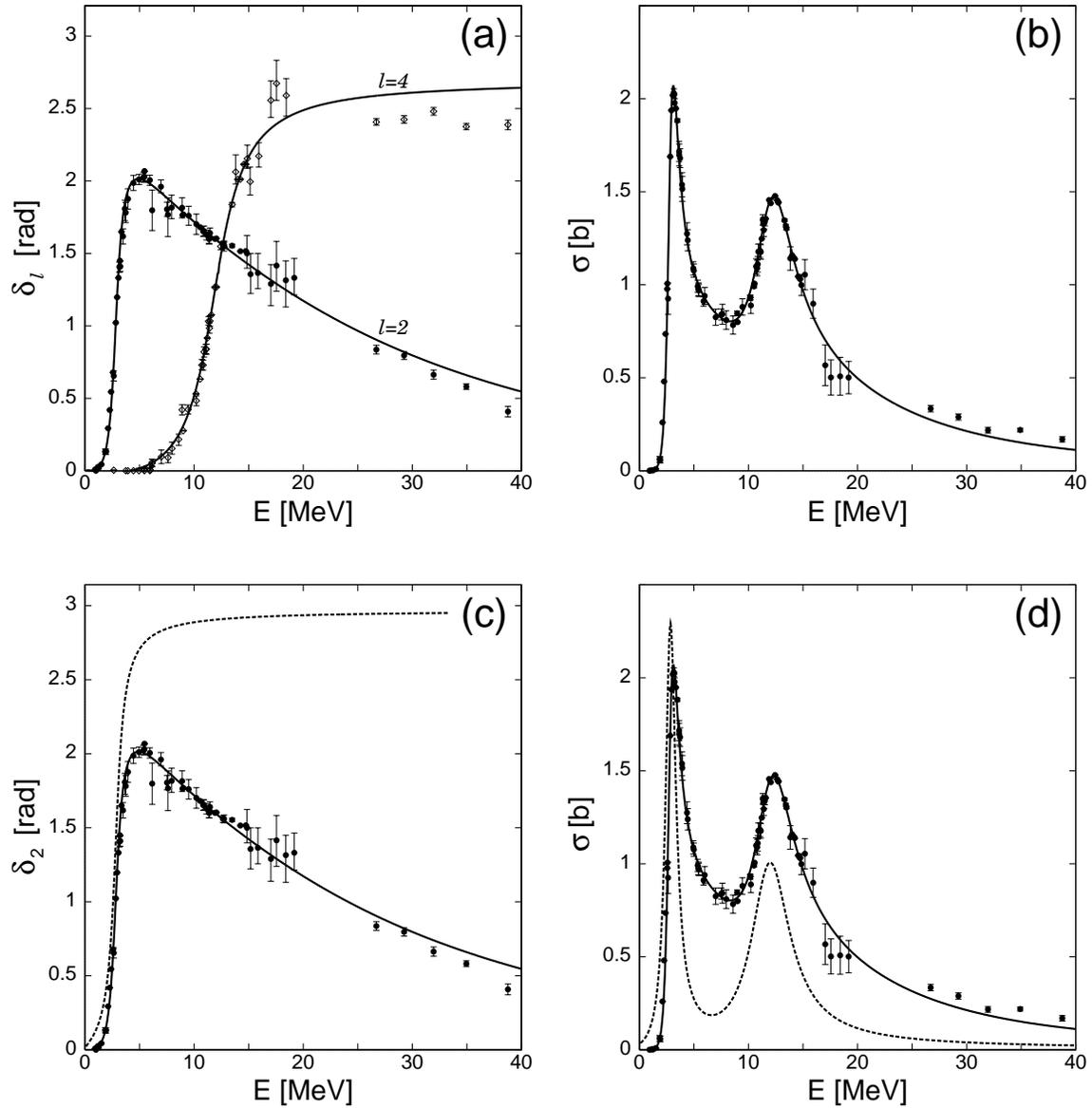}
\vspace*{7pt}
\caption{
\label{fig_3}
\small
$\alpha$--$\alpha$ elastic scattering.
(a) Experimental phase--shifts for the partial waves $\ell=2$, $\ell=4$, and
corresponding fits (solid lines) (see formula (\protect\ref{28})), vs. the
center of mass energy $E$. Experimental data are taken from Refs.
\protect\cite{Afzal,Buck,Chien,Darriulat,Tombrello}. The numerical values of
the fitting parameters are (see text): $I=0.76$ (MeV)$^{-1}$, $\alpha_0=1.6$,
$b_1=1.06\times 10^{-1}$ (MeV)$^{-1/2}$, $a_1=1.03$ (MeV)$^{-1/4}$, $g_0=0.72$,
$g_1=-7.5\times 10^{-3}$ (MeV)$^{-1}$, $g_2=2.0\times 10^{-5}$ (MeV)$^{-2}$,
$E_0=4.1$ MeV.
(b) Total cross--section computed by using the phase--shifts in (a) (see
formula (\protect\ref{29})).
(c) Phase--shift for the partial wave $\ell=2$. The solid line indicates the
phase--shift computed by using formula (\protect\ref{28}), which takes into
account both the resonance and antiresonance terms. The dashed line shows the
phase--shift computed by using only the resonance term (see formula
(\protect\ref{22})).
(d) Comparison between the total cross--section computed by accounting for both
the resonance and antiresonance terms (solid line), and that computed by using
only the resonance term (dashed line).
}
\end{figure}

In Fig. \ref{fig_3}a the fits of the experimental phase--shifts, obtained by means of 
equation (\ref{28}), for $\ell=2,4$ are shown, while Fig. \ref{fig_3}c shows the 
effect of the antiresonance on the $\ell=2$ phase--shift.
The resulting numerical values of the fitting parameters are given in the 
figure legend. It is clear that the phase--shifts obtained by using equation 
(\ref{28}) reproduce rather well the experimental data.
In particular, they are notably better than the ones obtained previously in 
Ref. \cite{Viano1}, where a hard--core model of the repulsive part of the 
interaction was implemented, and whose phase--shifts turned out to be not 
adequate for representing the nonresonant part of the phase--shifts over a 
sufficient energy range (see also Refs.\cite{Chien,Darriulat}). 
Table \ref{tab_1} summarizes the results of the analysis for what concerns 
the $2^+$ and $4^+$ resonances. 
The agreement with the experimental values appears quite good, but for a slight 
discrepancy in the $2^+$ total width $\Gamma$. However, as will be discussed 
later in this section, care must be taken in the interpretation of the resonance 
widths. It is worth remarking that some experimental indications of a 
$6^+$ state at $E_R \sim 28$ MeV and of an $8^+$ state at $E_R \sim 57$ MeV 
have been reported \cite{Ajzenberg}. Pushing forward our analysis, and computing 
$\delta_\ell$ from (\ref{28}) even for $\ell=6$ and $\ell=8$, we obtain 
a resonance $6^+$ at $E_R \sim 27.2$ MeV, and a $8^+$ resonance at 
$E_R \sim 47$ MeV. Then, with a single pair of pole trajectories (one for 
resonances and one for antiresonances) this rotational band of resonances can 
be fitted quite accurately.

\begin{table}[t]
\caption{\label{tab_1}\small $\alpha$--$\alpha$ elastic scattering. In the present 
work the resonance energy $E_R$, associated with the angular momentum $\ell$, 
is defined as the energy of the upward $\frac{\pi}{2}$--crossing of the 
corresponding phase--shift $\delta_\ell(E)$. The \textit{purely resonant} 
$\Gamma_R$ indicates the width of the resonance peak computed without the 
antiresonance contribution, while the \textit{total} $\Gamma$ stands for the width 
of the resonance peak accounting also for the antiresonance term.
$\frac{\Delta {\cal S}}{{\cal S}_\textrm{Res}}$ indicates the relative increase
of skewness of the resonance peak when the antiresonance contribution
is added to the pure resonant term; here ${\cal S}$ is evaluated by means of
${\cal S}_\textrm{phen}$.
}
\begin{center}
\begin{tabular}{ccccccc}\hline
$J^P$&$E_R$ [MeV] & $E_R$ [MeV] & $\Gamma_R$ [MeV] & $\Gamma$ [MeV] &
$\Gamma$ [MeV]    & $\frac{\Delta {\cal S}}{{\cal S}_\textrm{Res}}$ \\
 & (present work) & (Ref. \protect\cite{Ajzenberg}) & Purely resonant & Total &
(Ref. \protect\cite{Ajzenberg}) &   \\ \hline
$2^+$ & 3.23 & $3.27$ & $1.04$ & $2.58$ & $1.50$ & $20.53$ \\
$4^+$ & 12.6 & $11.6\pm0.3$ & $4.33$ & $4.91$ & $4.0 \pm 0.4$ & $2.35$  \\
\hline
\end{tabular}
\end{center}
\end{table}

From formula (\ref{29}) the total cross--section can be computed from the 
phase--shifts of Fig. \ref{fig_3}a: the result is shown in Fig. \ref{fig_3}b, 
while in Fig. \ref{fig_3}d the total cross--sections computed with and without 
the antiresonance term are compared. 

The analysis of the total cross--section arouses the issue regarding the 
definition of the resonance parameters, in particular of the width $\Gamma$ of 
the resonance. The extraction of the resonance parameters is model dependent, 
and many ways to proceed in practice have been presented in the literature 
(see, for instance, Refs. \cite{Cutkosky,Vrana} and the references therein).
The estimate of $\Gamma$ presents several difficult questions in the framework 
of the present theory as well as in the Breit--Wigner formalism. In both theories 
the main difficulty derives from the effect of the antiresonance which deforms 
the bell--shaped symmetry of the resonance peak. Further, we must note: \\
i) In the present theory, formula (\ref{23}) provides an estimate of the pure 
resonance width $\Gamma_R$, which must be understood as the width of the 
resonance in the absence of the antiresonance effect; as explained in section 
\ref{se:complex} an analogous estimate for the \textit{antiresonance width} 
$\Gamma_A$ is hardly definable. \\
ii) In the fit of the experimental cross--section within the Breit--Wigner 
theory, the estimate of $\Gamma$ is obtained by adding to the pure resonance a 
background term, which is supposed to be generated by the so--called 
\textit{potential scattering}. One obtains a purely phenomenological result.

\begin{figure}[th]
\centering
\includegraphics[scale=1.0]{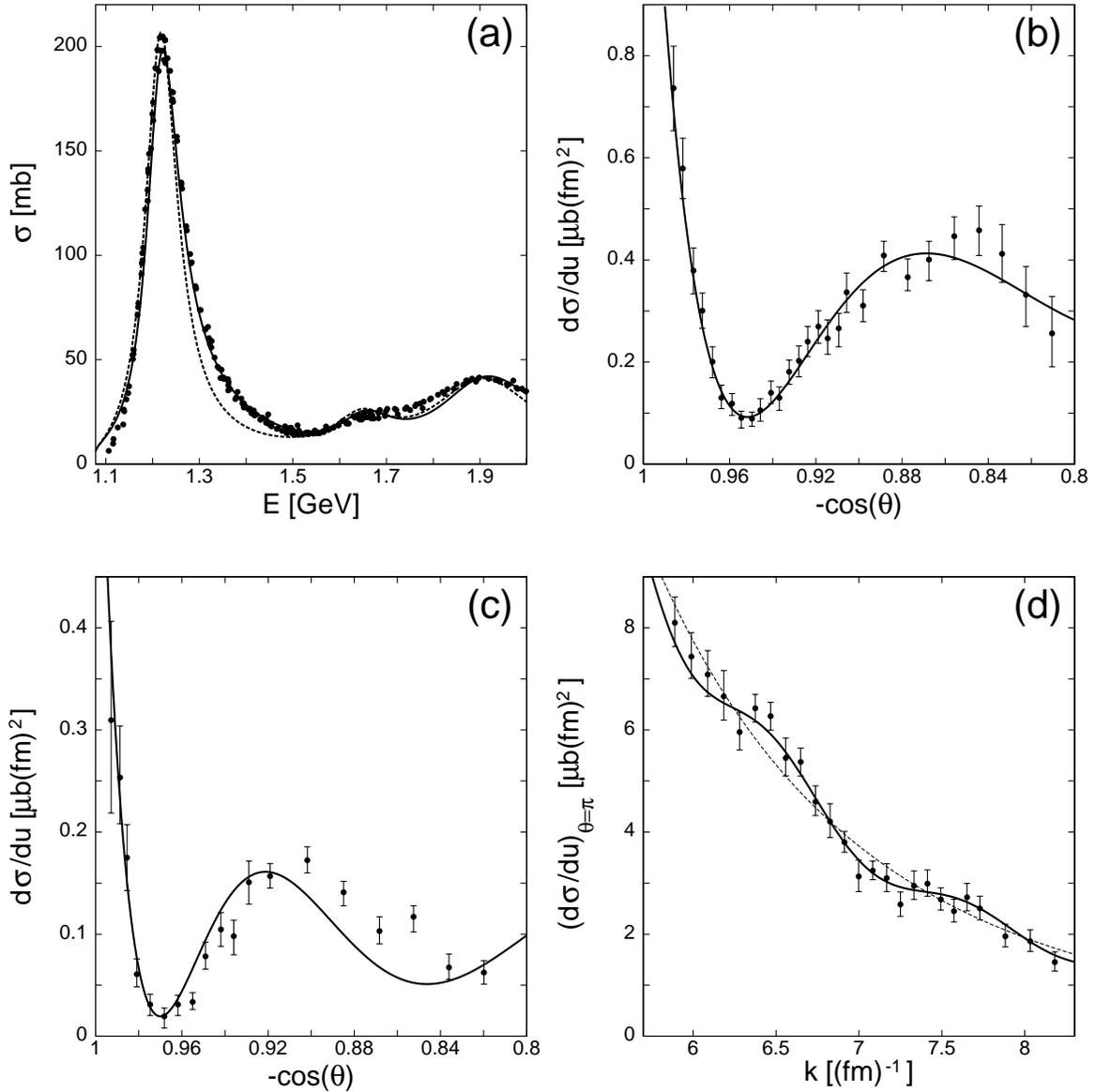}
\vspace*{7pt}
\caption{
\label{fig_4}
\small
$\pi^+$--p elastic scattering.
(a) Total cross--section. The experimental data (dots) are taken from
Ref. \protect\cite{Hagiwara}. The solid line indicates the total cross--section
computed by means of (\protect\ref{41}), and taking into account the
contributions of both the resonance and antiresonance poles generating
$\delta_\ell^{(+)}$ (formula (\protect\ref{42})), and the resonance pole
generating $\delta_\ell^{(-)}$.  The dashed line shows the total cross--section
computed by accounting only for the resonance poles generating
$\delta_\ell^{(+)}$ and $\delta_\ell^{(-)}$, respectively.
The fitting parameters are (see (\protect\ref{42}) through
(\protect\ref{44f})): $a_0^{(+)}=6.89\times 10^{-1}$,
$a_1^{(+)}=9.2\times 10^{-7}$ (MeV)$^{-2}$,
$b_1^{(+)}=9.0\times 10^{-5}$ (MeV)$^{-1}$,
$b_2^{(+)}=1.4\times 10^{-7}$ (MeV)$^{-2}$, $c_0=-0.5$,
$c_1=5.0\times 10^{-7}$ (MeV)$^{-2}$,
$g_0=2.0\times 10^{-6}$ (MeV)$^{-2}$, $g_1=3.0\times 10^{-12}$ (MeV)$^{-4}$,
$a_1^{(-)}=6.4\times 10^{-7}$ (MeV)$^{-2}$,
$b_1^{(-)}=1.25\times 10^{-4}$ (MeV)$^{-1}$.
(b) Differential cross--section vs $(-\cos\theta)$, at
$s = E^2 = 10.66$ (GeV)$^2$. The experimental data (dots) are taken from
Ref. \protect\cite{Baker}. The solid line shows the differential cross--section
computed by means of (\protect\ref{45}). The fitting parameters are:
$\Real\lambda=7.1$, $\Imag\lambda=1.3$, $B_0=1.61$ $\mu$b (fm)$^2$,
$B_1=8.0\times 10^{-4}$ $\mu$b (fm)$^2$.
(c) Differential cross--section vs $(-\cos\theta)$, at
$s = E^2 = 14.04$ (GeV)$^2$. The fitting parameters are: $\Real\lambda=9.2$,
$\Imag\lambda=1.2$, $B_0=0.778$ $\mu$b (fm)$^2$,
$B_1=1.6\times 10^{-5}$ $\mu$b (fm)$^2$.
(d) Differential cross--section at $\theta=\pi$ vs $k$. The experimental data
(dots) are taken from Ref. \protect\cite{Lennox}. The solid line shows the
differential cross--section computed by means of (\protect\ref{34}).
The fitting parameters are:
$C_0=1.85\times 10^6$ $\mu$b (fm)$^{2/3}$, $c=5.5$ (fm)$^{1/3}$,
$A=-9.3\times 10^{-2}$, $\omega=5.4$ fm, $\phi=0.84$.
The dashed line shows the differential cross--section computed by means of
formula (\protect\ref{33}), in which the oscillating term is absent.
}
\end{figure}

Reverting to our theory, in order to give a phenomenological estimate of the 
width $\Gamma$, two different situations should be distinguished:

I) The effect of the antiresonance is a small perturbation to the pure resonance 
(see, for instance, the leftmost resonance peak in Fig. \ref{fig_4}a in 
connection with the $\pi^+$--p elastic scattering). This means that the 
reference baseline of the pure resonance peak and that of the observed 
cross--section (comprising both resonance and antiresonance) coincide within 
a good approximation. 

\setlength{\unitlength}{1cm}

In this case we can proceed operatively as follows. First, from formula 
(\ref{23}) we obtain the estimate of $\Gamma_R$. Then, from the plot of the 
cross-section generated by only the pure resonant term, i.e. obtained by using 
only the pole singularity lying in the first quadrant of the $\lambda$--plane, 
we can recover the reference baseline of this almost symmetric bell--shaped 
distribution by equating its second central moment to the the value of 
$\Gamma_R$ (evaluated by means of (\ref{23})). Next, keeping fixed this 
baseline, we evaluate the second central moment of the distribution which fits 
the experimental cross--section peak (i.e., accounting also for the antiresonant 
term). We can take as an estimate of the total width $\Gamma$ the value of this 
second moment. As a measure of the degree of asymmetry of the resonance peak,
which can be ultimately ascribed to the composite structure (non--elementariness) 
of the interacting particles, we take the statistical skewness 
${\cal S}_\textrm{stat}$ of the distribution, defined as 
${\cal S}_\textrm{stat} = \frac{\mu_3}{\mu_2^{3/2}}$, where $\mu_2$ and $\mu_3$ are 
respectively the second and third central moments of the distribution (see 
Table \ref{tab_2} for the numerical values related to the $\pi^+$--p elastic 
scattering).
This procedure works reasonably well as far as the asymmetry of the resonance 
peak is not too large and can be regarded as a small perturbation of the pure 
resonance effect, as in the case of the $\Delta(\frac{3}{2},\frac{3}{2})$ 
resonance in the $\pi^+$--p elastic scattering, which will be treated in the 
next subsection.

II) The asymmetry of the bell--shaped peak is very large, and the antiresonance 
effect cannot be regarded as a small perturbation to the pure resonance (see 
figs. \ref{fig_3}b and \ref{fig_3}d in connection with the $\alpha$--$\alpha$ elastic scattering). 
In this case the increase of the cross--section corresponding to the downward 
crossing of $\frac{\pi}{2}$ of the phase--shift is relevant and deforms 
considerably the shape of the pure resonant peak. Moreover, the reference 
baseline of the pure resonance peak and that of the observed experimental 
cross--section differ significantly, so that a \textit{statistical} approach as 
the one described in the previous case (I) cannot be adopted. We are forced 
to follow a more pragmatic attitude. Since the asymmetry due to the 
antiresonance effect sets in just after the resonance maximum, we can regard
$\Gamma$ as the full--width at half--maximum of the resonance peak, like in 
the Breit--Wigner theory. From the plot of the pure resonance cross--section 
we obtain a bell--shaped distribution whose full--width at half--maximum agrees 
with the value of $\Gamma_R$ evaluated by formula (\ref{23}). Then we can 
give an estimate of the total width $\Gamma$ by evaluating the full--width at 
half--maximum of the curve fitting the experimental cross--section (i.e., 
including both resonance and antiresonance terms). 
In this case the asymmetry of the resonance peak can be estimated by using a 
\textit{phenomenological skewness} ${\cal S}_\textrm{phen}$, defined as follows: 
first we compute the difference between the two half--maximum semi--widths, 
measured with respect to the energy of resonance $E_R$; then the degree of 
asymmetry ${\cal S}_\textrm{phen}$ is defined as the ratio between this value 
and the full--width $\Gamma$ (see Table \ref{tab_1} for numerical values 
related to the $\alpha$--$\alpha$ elastic scattering).

\subsection{Resonances and surface waves in \boldmath{$\pi^+$}--p elastic scattering}
\label{subse:pionproton}
In the analysis of the $\pi^+$--p scattering the spin of the proton must be 
taken into account.  Therefore, we start recalling rapidly the main formulae 
for the scattering amplitude in the case of spin--$0$--spin--$\frac{1}{2}$ 
collision. In particular, we have the spin--non--flip amplitude and the 
spin--flip amplitude, which respectively read:
\begin{subequations}
\label{36}
\begin{eqnarray}
f(k,\theta)&=&\frac{1}{2\rmi k}\sum_{\ell=0}^\infty
\left[(\ell+1)(\SSSS_\ell^{(+)}-1)+
\ell\,(\SSSS_\ell^{(-)}-1)\right]\,P_\ell(\cos\theta), \label{36a} \\
g(k,\theta)&=&\frac{1}{2k}\sum_{\ell=0}^\infty
(\SSSS_\ell^{(+)}-\SSSS_\ell^{(-)})\,P^{(1)}_\ell(\cos\theta), \label{36b}
\end{eqnarray}
\end{subequations}
where $P^{(1)}_\ell(\cos\theta)$ is the associated Legendre function, and 
\begin{subequations}
\label{38}
\begin{eqnarray}
\SSSS_\ell^{(+)} = e^{2\rmi\delta_\ell^{(+)}}, \label{38a} \\
\SSSS_\ell^{(-)} = e^{2\rmi\delta_\ell^{(-)}}, \label{38b}
\end{eqnarray}
\end{subequations}
$\delta_\ell^{(\pm)}$ being the phase--shift associated with the partial wave 
with total angular momentum $j=\ell\pm\frac{1}{2}$. The differential 
cross--section is given by
\beq
\frac{\rmd\sigma}{\rmd\Omega}= |f|^2 + |g|^2,
\label{40}
\eeq
if the proton target is unpolarized, and if the Coulomb scattering is neglected, 
as it is admissible at energy sufficiently high. Let us note that the Sommerfeld 
parameter $\eta = \frac{e^2}{\hbar v}$ at $E \simeq 1200$ MeV (close to the 
energy of the $\Delta(\frac{3}{2},\frac{3}{2})$ resonance) is of the order of 
$0.04$. Next, integrating over the angles and taking into account the 
orthogonality of the spherical harmonics, we obtain for the total cross--section 
the following expression:
\beq
\sigma_\textrm{tot} = \frac{2\pi}{k^2} \sum_{j,\ell} (2j+1) \sin^2\delta_{\ell,j},
\label{41}
\eeq
where $j=\ell\pm\frac{1}{2}$, $\delta_{\ell,j}=\delta_{\ell,\ell\pm 1/2}$, 
$\delta_{\ell,\ell+1/2}\equiv\delta_{\ell}^{(+)}$, 
$\delta_{\ell,\ell-1/2}\equiv\delta_{\ell}^{(-)}$.

We put at the center of our analysis the $\Delta(\frac{3}{2},\frac{3}{2})$ 
resonance. It is considered the first member of a family of resonances whose 
$J^P$ values are precisely given by: $\frac{3}{2}^{+}$, $\frac{7}{2}^{+}$, 
$\frac{11}{2}^{+}$, $\frac{15}{2}^{+}$, $\frac{19}{2}^{+}$.
It has been suggested \cite{Dalitz} that this sequence could correspond to an 
even rotational band of the proton states whose angular momentum is given by: 
$L=0^+,2^+,4^+,6^+,8^+$. We could as well have an odd rotational band of proton 
states with angular momentum: $L=1^-,3^-,\ldots$. But, in the $\pi^+$--p elastic 
scattering, we observe only one resonance with $J^P =\frac{1}{2}^{-}$, which 
could correspond to the first member (i.e., $L=1^-$) of this odd rotational 
band\footnote{We have already treated in I the 
generation of even and odd rotational bands starting from the dynamics of the 
three--body problem. We intend to present elsewhere a more detailed analysis of 
this type of bands in the specific case of confining potentials acting among 
three quarks.} (see Ref. \cite{Dalitz}).

Let us now focus on the first family of resonances. In our model they should be 
fitted by the trajectory of one pole lying in the first quadrant of the complex 
angular momentum plane. But, since the proton and the pion are composite 
particles, the antiresonances should play a role. Accordingly, we should add 
the contribution of a pole in the fourth quadrant of the complex angular
momentum plane. Furthermore, since the spin of the proton is fixed, we limit 
ourselves to perform the analytic continuation of the partial waves from 
integers to complex values of the angular momentum $\ell$. Therefore, we shall 
fit the resonances belonging to the family whose first member is 
$\Delta(\frac{3}{2},\frac{3}{2})$, by writing for 
$\delta_{\ell,\ell+1/2}\equiv\delta_{\ell}^{(+)}$ the following expression:
\begin{eqnarray}
\label{42}
\delta_{\ell}^{(+)}&=&\sin^{-1}\left\{\frac{1-(-1)^\ell}{2}
\frac{\beta_R^{(+)}(2\alpha_R^{(+)}+1)}
{\left\{\left[(\ell-\alpha_R^{(+)})^2+
(\beta_R^{(+)})^2\right]\left[(\ell+\alpha_R^{(+)}+1)^2+(\beta_R^{(+)})^2\right]
\right\}^{1/2}}\right\} \nonumber \\
&+& \sin^{-1}\left\{\frac{1-(-1)^\ell}{2}
\frac{-\beta_A^{(+)}(2\alpha_A^{(+)}+1)}
{\left\{\left[(\ell-\alpha_A^{(+)})^2+(\beta_A^{(+)})^2\right]
\left[(\ell+\alpha_A^{(+)}+1)^2+(\beta_A^{(+)})^2\right]\right\}^{1/2}}\right\}.
\end{eqnarray}
(Let us note that the factor $\frac{1-(-1)^\ell}{2}$ in formula (\ref{42}) 
and in the next formula (\ref{43}), instead of $\frac{1+(-1)^\ell}{2}$, 
is due to the fact that we interpolate odd values of $\ell$).

We can now pass to consider the second family of resonances, whose first member 
is $\Delta(\frac{1}{2},\frac{3}{2})$. But, as mentioned above, this resonance is 
the sole member of this sequence which is phenomenologically observed in elastic 
scattering. We thus introduce a second trajectory of a pole lying in the first 
quadrant. In view of the small effect of this resonance on the total 
cross--section, we neglect the corresponding antiresonance pole. 
Then, with obvious meaning of the notations, we write for 
$\delta_{\ell,\ell-1/2}\equiv\delta_{\ell}^{(-)}$:
\beq
\delta_{\ell}^{(-)}=\sin^{-1}\left\{\frac{1-(-1)^\ell}{2} 
\frac{\beta_R^{(-)}(2\alpha_R^{(-)}+1)}
{\left\{\left[(\ell-\alpha_R^{(-)})^2+(\beta_R^{(-)})^2\right]
\left[(\ell+\alpha_R^{(-)}+1)^2+(\beta_R^{(-)})^2\right]
\right\}^{1/2}}\right\}.
\label{43}
\eeq
We then fit the total cross--section parametrizing $\alpha_R^{(\pm)}$, 
$\beta_R^{(\pm)}$, $\alpha_A^{(+)}$ and $\beta_A^{(+)}$ as follows:
\begin{subequations}
\label{44}
\begin{eqnarray}
\alpha_R^{(+)} &=& a_0^{(+)} + a_1^{(+)} (E^2 - E_0^2), \label{44a} \\
\beta_R^{(+)} &=& b_1^{(+)} \sqrt{E^2 - E_0^2} + b_2^{(+)} (E^2 - E_0^2), 
\label{44b}\\
\alpha_A^{(+)} &=& c_0 + c_1 (E^2 - E_0^2), \label{44c} \\
\beta_A^{(+)} &=& g_0 (E^2 - E_0^2) + g_1 (E^2 - E_0^2)^2, \label{44d} \\
\alpha_R^{(-)} &=& a_1^{(-)} (E^2 - E_0^2), \label{44e} \\
\beta_R^{(-)} &=& b_1^{(-)} \sqrt{E^2 - E_0^2}, \label{44f}
\end{eqnarray}
\end{subequations}
where $E$ is the energy in the center of mass frame, and $E_0$ is the rest mass 
of the $\pi^+$--p system.

Substituting the values $\delta_\ell^{(\pm)}$ (formulae (\ref{42}) through 
(\ref{44f})) in formula (\ref{41}) we can fit the total cross--section (the data 
are taken from Ref. \cite{Hagiwara}). The result is shown in Fig. \ref{fig_4}a, 
where the total cross--sections computed with (solid line) and without (dashed 
line) the antiresonance term are compared (see the figure legend for numerical 
details). The fit is very satisfactory, and shows with clear evidence the effect 
of the antiresonance corresponding to the resonance 
$\Delta(\frac{3}{2},\frac{3}{2})$. The difference between the two curves reveals 
the composite structure of the interacting particles.

\begin{table}[t]
\caption{
\label{tab_2}
\small
$\pi^+$--p elastic scattering. Notice that, in the rightmost column,
${\cal S}$ is evaluated by means of ${\cal S}_\textrm{stat}$.
For the other definitions, see the legend of Table \protect\ref{tab_1}.
}
\begin{center}
\begin{tabular}{ccccccc}\hline
$J^P$ & Mass [MeV] & Mass [MeV] & $\Gamma_R$ [MeV] & $\Gamma$ [MeV] &
$\Gamma$ [MeV] & $\frac{\Delta{\cal S}}{{\cal S}_\textrm{Res}}$ \\
 & (present work) & (Ref. \protect\cite{Hagiwara}) & Purely resonant & Total &
(Ref. \protect\cite{Hagiwara}) & \\ \hline
$\frac{3}{2}^+$&1231&$1230 \div 1234$ & $84$ & 117 & $115 \div 125$ & 4.40 \\
$\frac{7}{2}^+$&1941&$1940 \div 1960$ & $273$ & 308 & $290 \div 350$ & 3.80 \\
$\frac{11}{2}^+$&2445&$2300 \div 2500$ & $374$ & 410 & $300 \div 500$ & 3.24 \\
\hline
$\frac{1}{2}^-$&1655&$1615 \div 1675$ & $147$ & --- & $120 \div 180$ & ---  \\
\hline
\end{tabular}
\end{center}
\end{table}

In Table \ref{tab_2} the analysis for family of resonances 
$\Delta(\frac{3}{2},\frac{3}{2})$, $\Delta(\frac{7}{2},\frac{3}{2})$, 
$\Delta(\frac{11}{2},\frac{3}{2})$, $\Delta(\frac{1}{2},\frac{3}{2})$
is summarized.
In particular we give the energy location, the purely resonant and total widths, 
and the skewness ascribable to the antiresonance phenomenon. It should be 
remarked that the values of the $\Delta(\frac{11}{2},\frac{3}{2})$ resonance, 
which is not visible in Fig. \ref{fig_4}a, have been extrapolated by computing 
$\delta_{\ell=5}^{(+)}$ with the parameters obtained from the analysis of the 
$\Delta(\frac{3}{2},\frac{3}{2})$ and $\Delta(\frac{7}{2},\frac{3}{2})$ 
resonances (see the legend of Fig. \ref{fig_4}). It is worth noticing from the 
last column in Table \ref{tab_2} that the degree of asymmetry associated with 
the resonance peaks decreases notably when the angular momentum $\ell$ increases. 
This behaviour was expected since the asymmetry of the resonance peaks is due to 
the antiresonances, whose effect tends to disappear as $\ell$ increases (see also 
Table \ref{tab_1} for a similar behaviour in connection with the 
$\alpha$--$\alpha$ scattering).

As shown in Fig. \ref{fig_4}a we can fit the total cross--section up to a value 
of $E$ of the order of $2000$ MeV. At higher energy, the elastic unitarity 
condition is largely violated, and the fitting formula should be modified 
accordingly. 
Furthermore, the resonances $\frac{15}{2}^{+}$ and $\frac{19}{2}^{+}$ do not 
display sharp peaks in the total cross--section. This means that the resonances
evolve into surface waves in the sense described in section \ref{se:from}. 
At these energies the partial--wave analysis cannot be properly applied, nor 
it has meaning to separate the $\delta_\ell^{(+)}$ from the $\delta_\ell^{(-)}$ 
trajectories. As explained in section \ref{se:from} we can try two different 
types of fits: \textit{i}) at fixed energy; \textit{ii}) at fixed angle: 
i.e., $\theta=\pi$. We start with the first type of fit.  With this in mind 
we approximate the differential cross--section at backwards with the following 
formula:
\beq
\frac{\pi}{k^2}\frac{\rmd\sigma}{\rmd\Omega}\simeq
B_0 |P_\lambda(-\cos\theta)|^2 + B_1 |P_\lambda^{(1)}(-\cos\theta)|^2.
\label{45}
\eeq
In formula (\ref{45}) we introduce the term 
$B_1 |P_\lambda^{(1)}(-\cos\theta)|^2$, which gives the contribution to the 
differential cross--section of the spin--flip amplitude (see formulae
(\ref{36})). In figs. \ref{fig_4}b and \ref{fig_4}c we present two fits of the 
differential cross--section in the backward angular region $0.8<-\cos\theta<1.0$, 
at fixed energy. The fits are satisfactory, and $B_1$ turns out to be negligible, 
compared to $B_0$. Let us moreover note that the values of $\Real\lambda$
obtained by these fits indicate that we do not have resonances at 
$J^P=\frac{15}{2}^{+}$ and $J^P=\frac{19}{2}^{+}$, but backward peaks due to 
creeping wave effects, which can be described by the Sommerfeld poles, instead 
of by the Regge's ones.

Finally, in Fig. \ref{fig_4}d we present a fit at fixed angle: $\theta=\pi$. 
The fit is performed by means of formula (\ref{34}), derived in section 
\ref{se:from}. 
It presents a clear evidence of an oscillating pattern due to the interference 
of grazing rays which undergo a different number of shortcuts. As the energy
increases, the radius of the central core increases too, and the shortcuts are 
progressively suppressed: the oscillating pattern is damped.

\end{document}